\documentclass[letterpaper,twocolumn,prl,aps,superscriptaddress,floatfix,showpacs]{revtex4-1}

\usepackage{multirow,amssymb,amsbsy,amsmath}
\usepackage{graphicx}
\usepackage{verbatim}
\usepackage{booktabs}
\usepackage{dcolumn}
\makeatletter
\def\bra#1{\mathinner{\langle{#1}|}}
\def\ket#1{\mathinner{|{#1}\rangle}}

\usepackage{color}

\begin{document}

\title{Nonlocal memory assisted entanglement distribution in optical fibers}

\author{Guo-Yong Xiang$^{\ddag}$}
\affiliation{Key Laboratory of Quantum Information, University of Science and 
Technology of China, CAS, Hefei, 230026, China}
\email{These authors contributed equally to this work.}

\author{Zhi-Bo Hou$^{\ddag}$}
\affiliation{Key Laboratory of Quantum Information, University of Science and 
Technology of China, CAS, Hefei, 230026, China}

\author{Chuan-Feng Li}
\email{cfli@ustc.edu.cn}
\affiliation{Key Laboratory of Quantum Information, University of Science and 
Technology of China, CAS, Hefei, 230026, China}

\author{Guang-Can Guo}
\affiliation{Key Laboratory of Quantum Information, University of Science and
Technology of China, CAS, Hefei, 230026, China}

\author{Heinz-Peter Breuer}
\affiliation{Physikalisches Institut, Universit\"at Freiburg,
Hermann-Herder-Strasse 3, D-79104 Freiburg, Germany}

\author{Elsi-Mari Laine}
\affiliation{QCD Labs, COMP Centre of Excellence, Department of Applied Physics, Aalto University, P.O. Box 13500, FI-00076 AALTO, Finland}
\affiliation{Turku Centre for Quantum Physics, Department of Physics and 
Astronomy, University of Turku, FI-20014 Turun yliopisto, Finland}

\author{Jyrki Piilo}
\email{jyrki.piilo@utu.fi}
\affiliation{Turku Centre for Quantum Physics, Department of Physics and 
Astronomy, University of Turku, FI-20014 Turun yliopisto, Finland}

\date{\today}

\begin{abstract}

Successful implementation of several quantum information and communication 
protocols require distributing entangled pairs of quantum bits in reliable 
manner. While there exists a substantial amount of recent 
theoretical and experimental activities dealing with non-Markovian quantum 
dynamics, experimental application and verification of the usefulness of 
memory-effects for quantum information tasks is still missing. We combine 
these two aspects and show experimentally that a recently introduced concept of 
nonlocal memory effects allows to protect and distribute polarization 
entangled pairs of photons in efficient manner within polarization-maintaining (PM) optical fibers. The introduced 
scheme is based on correlating the environments, 
i.e.~frequencies of the 
polarization entangled photons, 
before their physical distribution. When comparing 
to the case without nonlocal memory effects, we demonstrate at least 12-fold 
improvement in the channel, or fiber length, for preserving the highly-entangled 
initial polarization states of photons against dephasing.  
\end{abstract}


\maketitle

Quantum information processing holds the promise of faster computation of specific tasks and more secure communication than classically achievable~\cite{Galindo02,NC}. Therefore, during the last two decades, a wide community of researchers have made substantial efforts to convert the ideas of quantum information theory into practice and to realize powerful quantum information enabled devices. 
Considering the fragility of quantum states and their properties, remarkable advances have been accomplished on various physical platforms~\cite{qc1,qc2,qc3}. Despite of this progress, however, large scale quantum computers are yet to be seen and pragmatic quantum communication applications still face considerable challenges. One of the key resources for quantum computing and communication is the ability to generate and distribute entanglement. However, the sharing of quantum correlations is very challenging task, since entanglement is vulnerable to environmental noise, and there is a need to develop novel methods for this purpose. Our proposal and its verification is based on the observation that noise can be fought with noise, when considering open quantum systems.

We focus on optical realizations of entanglement generation and distribution, i.e.~parametric 
downconversion~\cite{kwiat00} to generate polarization entangled photon pairs and their distribution with PM optical fibers. While there exists recent impressive experiments to distribute entangled photons over long distances in free space~\cite{yin12,ma12}, we concentrate on fiber-based distribution schemes due to their flexibility in terms of distribution locations and relevance for the construction of quantum networks. 
Moreover, our scheme distributes each photon in its own PM-fiber and does not require post-distribution compensation as in some earlier studies using a single PM-fiber with auxiliary photon~\cite{yama08} or single-mode and non-zero dispersion shifted fiber~\cite{zei07}.
 
Decoherence is generally dealt within the theoretical framework of open quantum systems~\cite{BP} and non-Markovian quantum dynamics has been studied in great deal in recent
years. There has been rapid progress both in
theory~\cite{nm3,nm4,nm5,nm6,nm7,nm10,nm11,nlnm} and experimental 
control~\cite{nm12,nm13,nm14,nm15,nm16,nm17}, and the first theoretical proposals for exploiting non-Markovianity for quantum information processing and metrology exist~\cite{nmqi1,nmqi2,nmqi3}. However, an experimental demonstration of the exploitation of non-Markovianity in quantum information and entanglement protection is still missing.

It was recently shown theoretically that initial correlations in the composite environment of a bipartite quantum system can generate memory effects in the open system dynamics \cite{nlnm}. These memory effects are nonlocal since the constituent parts of the open system undergo a  memoryless decoherence process and each subsystem alone is not able to detect the presence of global non-Markovianity. 
We demonstrate how the nonlocal effects can be exploited to make the bipartite open system to behave effectively as a closed 
one with subsequent preservation of coherences and entanglement between two quantum bits.
We show experimentally that increasing the initial correlations between the frequencies of downcoverted photons and nonlocal memory effects, allow the distribution of entanglement over at least twelve times longer distances than in the absence of correlations.

In general, for multipartite quantum systems, decoherence gives rise to the decay of entanglement and, in certain cases, entanglement sudden death. 
If the noise experienced by the multipartite open system is composed of local memoryless noise channels $\Phi_n$ causing
decoherence, and the total influence is given by the tensor product of the local channels, i.e. $\Phi=\otimes_n \Phi_n$, entanglement of the multipartite state will be lost. However, when there are initial correlations between the local environments, $\Phi$ is not equal to the tensor product of the local memoryless channels, $\Phi\neq\otimes_n \Phi_n$, and this leads to the possibility to protect the global properties of the multipartite system even if the constituent parts are subjected to decoherence~\cite{nlnm}. 

A representative example of a nonlocal decoherence process is polarization mode dispersion (PMD) for polarization entangled photons created in spontaneous parametric downconversion and distributed in optical fibers. The parametric downconversion process allows the creation of correlations between the frequencies of the photons, and optical fibers couple locally the frequency and the polarization degree of freedom of the photons, causing noise.
The frequency correlations therefore give rise to correlated noise, which, as we will see later, allows robust entanglement protection. 

A general PMD channel for two qubits distributed via optical fibers of lengths $L_1$ and $L_2$, respectively, can be written as
\begin{equation}
\Phi_{L_1, L_2} (\ket{ij}\bra{kl})=f_{ijkl}(L_1,L_2)\ket{ij}\bra{kl},
\end{equation}
with $i,j,k,l\in \left\{H,V\right\}$, $f_{klij}=f^*_{ijkl}$, and
\begin{equation}\label{eq1b}
 f_{ijkl} = \left\{ \begin{array}{ll}
 1 & {\mbox{if}} \;\;\; i=k,j=l \\
 G(0,\tau_2) & {\mbox{if}} \;\;\; i=k,j=H,l=V \\
 G(\tau_1,0) & {\mbox{if}} \;\;\; i=H,k=V,j=l \\
 G(\tau_1,\tau_2) & {\mbox{if}} \;\;\; i=j=H, k=l=V\\
 G(\tau_1,-\tau_2) & {\mbox{if}} \;\;\; i=l=H, j=k=V,\\
 \end{array}\right. \nonumber
\end{equation}
where
\begin{equation}
G(\tau_1,\tau_2) = \int d\omega_1 \int d\omega_2 P(\omega_1,\omega_2) e^{-i(\omega_1\tau_1+\omega_2\tau_2)}.
\end{equation}
Here, $H$ ($V$) denotes the horizontal (vertical) polarization direction of the 
photon, $P(\omega_1,\omega_2)$ is the joint frequency distribution of the two 
photons, and $\tau_m=\Delta n_m L_m/c$ with $\Delta n_m$ the 
birefringences of the two fibers ($m=1,2$). 

\begin{figure}[tb]
\centering
\includegraphics[width=0.4\textwidth]{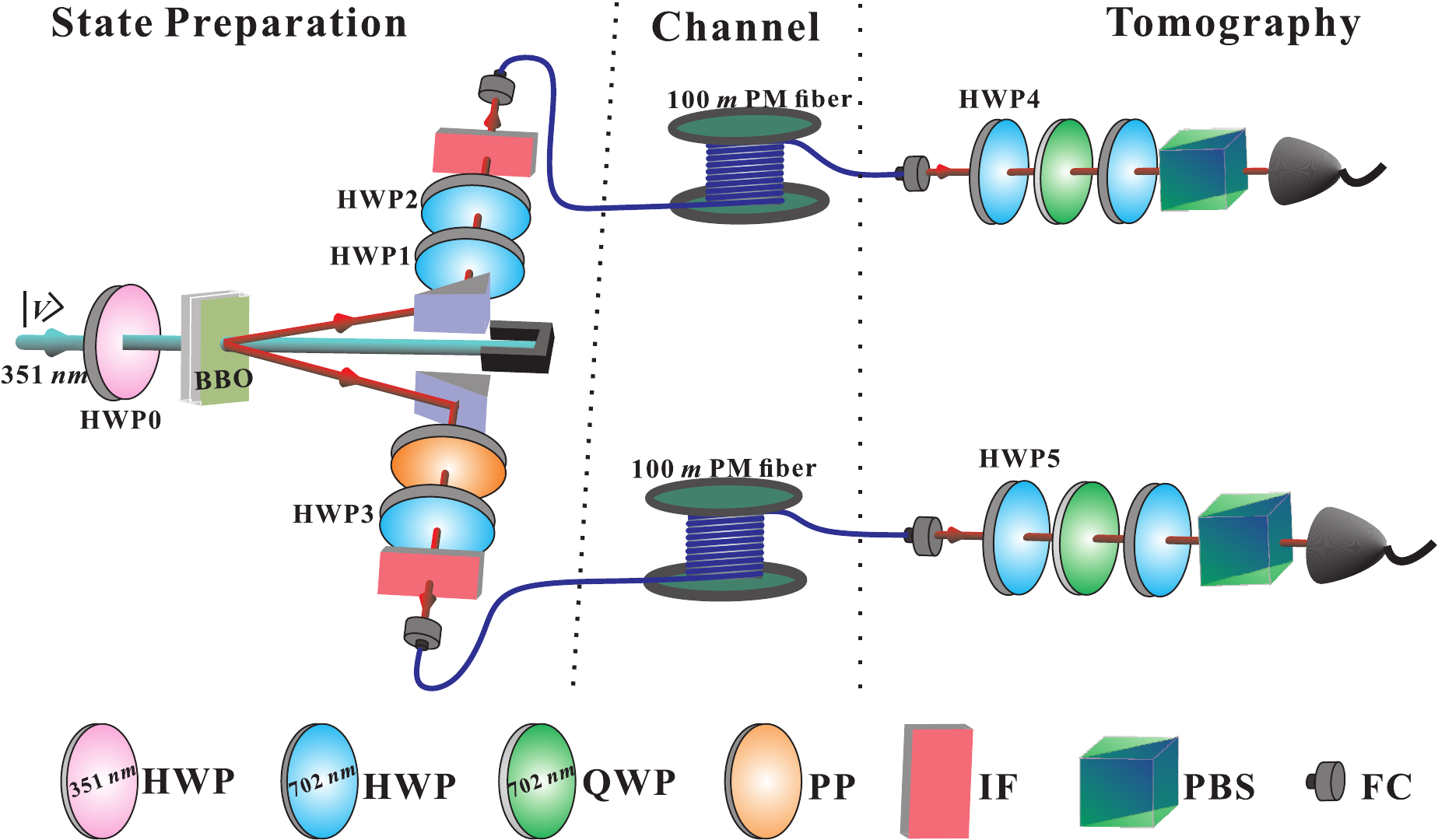}
\caption{\label{fig1} 
(color online) The experimental setup.
The apparatus consists of three parts: i) polarization state preparation of the two-photon state and the control of frequency correlations ii) PMD quantum channel by two $100$m long PM-optical fibers iii) state tomography. 
Key to the components: HWP--half-wave plate for 351.1 nm (pink) and 702.2 nm (blue), BBO -- $\beta$-barium borate crystal, QWP--quarter-wave plate, PP--phase plate, IF--interference filter, PBS--polarization beam splitter, and FC--fiber connector. 
}
\end{figure}

If the initial two-photon frequency distribution $P(\omega_1,\omega_2)$ does not factorize, the combined channel $\Phi_{L_1,L_2}$ for the two photons is nonlocal, i.e., $\Phi_{L_1,L_2}\neq \Phi_{L_1}\otimes \Phi_{L_2}$. Since we are interested in the capability of distributing entanglement, let us concentrate on how the PMD channel influences the Bell state $\ket{\Psi}=\frac{1}{\sqrt{2}}(\ket{HH}+\ket{VV})$. After passing the PMD channel, the state becomes
\begin{eqnarray}
\Phi_{L_1, L_2}(\ket{\Psi}\bra{\Psi})&=& \frac{1}{2}\Big[
\ket{HH}\bra{HH}+\ket{VV}\bra{VV}\nonumber\\
&+&G\left(\Delta n_1 L_1/c,\Delta n_2 L_2/c\right) \ket{HH}\bra{VV}\nonumber\\
&+&G^*\left(\Delta n_1 L_1/c,\Delta n_2 L_2/c\right) \ket{VV}\bra{HH} \Big] .\nonumber \\
\end{eqnarray}
The initial entanglement, measured by the concurrence, is equal to one and after 
the channel the entanglement is given by 
$E(L_1,L_2)=|G(\Delta n_1 L_1/c,\Delta n_2 L_2/c)|$.

We assume that the two-photon frequency distribution has a Gaussian form
\begin{equation} \label{GAUSS}
P(\omega_1,\omega_2) = \frac{1}{2\pi\sqrt{\textrm{det}~C}}
 e^{-\frac{1}{2}
 \left(\vec{\omega}-\vec{\langle \omega \rangle }\right)^TC^{-1}
 \left(\vec{\omega}-\vec{\langle \omega \rangle }\right)},
\end{equation}
where $\vec{\omega}=(\omega_1,\omega_2)^T$, $\vec{\langle \omega \rangle }=(\langle \omega_1 \rangle,\langle \omega_2 \rangle)^T$, and $C=(C_{ij})$ is given by $C_{ij}=\langle\omega_i\omega_j\rangle-\langle\omega_i\rangle\langle\omega_j\rangle$ with  $\langle \omega_1 \rangle = \langle \omega_2 \rangle =\omega_0/2$, $C_{11} = C_{22} = \langle\omega^2_i\rangle-\langle\omega_i\rangle^2$.
Therefore, the decoherence function is given by
\begin{equation} 
 G(\tau_1,\tau_2)=\exp\left[ \frac{i\omega_0}{2}(\tau_1+\tau_2)
 -\frac{C_{11}}{2}\left(\tau_1^2 + \tau_2^2
 + 2K\tau_1\tau_2\right)\right],
\end{equation}
where $K=C_{12}/\sqrt{C_{11}C_{22}}=C_{12}/C_{11}$ is the 
correlation coefficient between the two frequencies. 
Consequently, the entanglement after the action of the channel reads
\begin{equation}
E(L_1,L_2)=e^{-\frac{C_{11}}{2}\left[\left(\frac{\Delta n_1 L_1}{c}\right)^2+\left(\frac{\Delta n_2 L_2}{c}\right)^2+\frac{2 K \Delta n_1\Delta n_2 L_1 L_2}{c^2}\right] }.
\end{equation}
Clearly, if $K=-1$ corresponding to strong nonlocal memory effects~\cite{nlnm} and $\Delta n_1 L_1=\Delta n_2 L_2$, then for any fiber length, 
the entanglement is frozen and always equal to one. Thus, perfect anticorrelations 
between the frequencies allow entanglement distribution over arbitrary distances 
in the presence of PMD noise.

\begin{figure}[tb]
\centering
\includegraphics[width=0.45\textwidth]{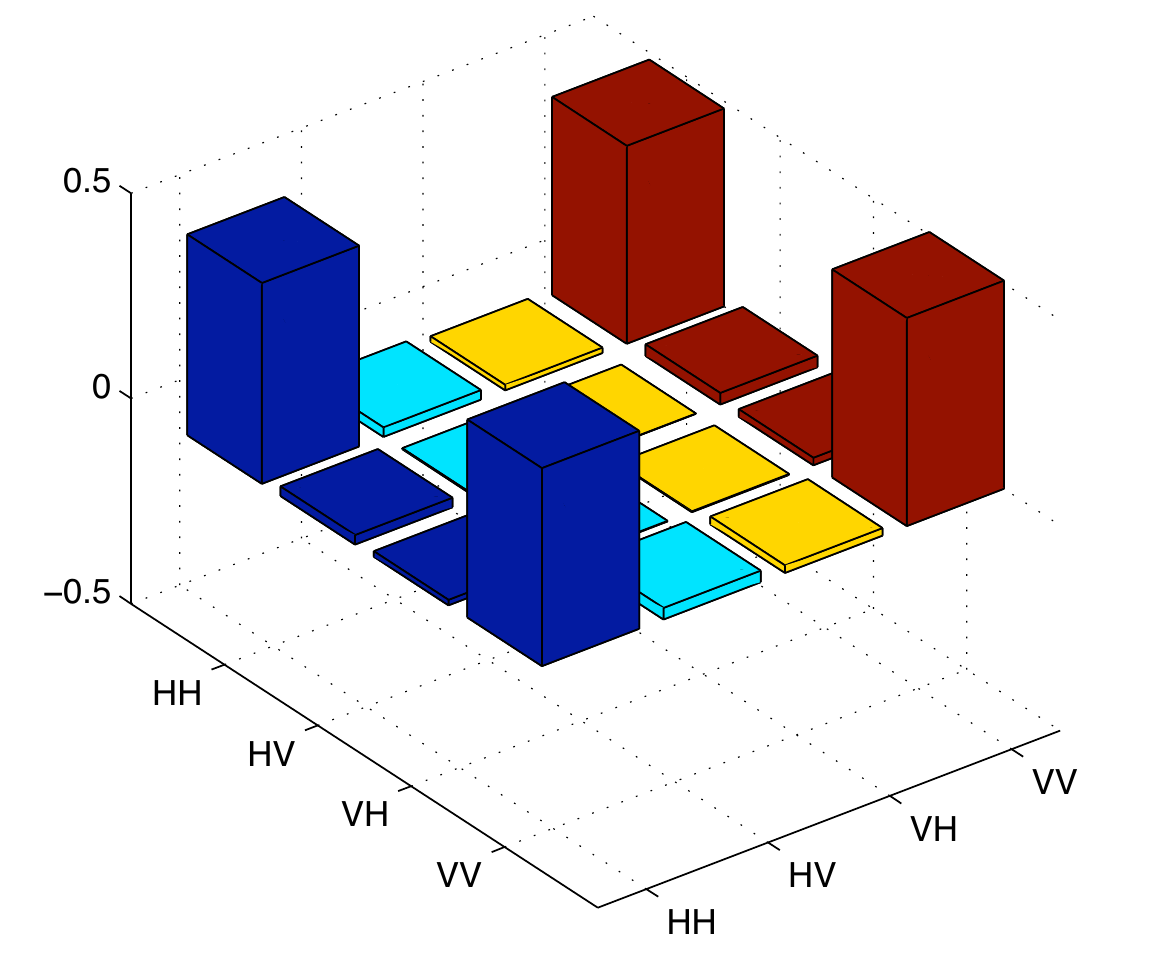}
\caption{\label{fig2} 
(color online) Tomography of the initial polarization entangled two-photon state.
For highly anticorrelated frequencies of the two photons, the tomographic polarization state density
matrix, shown here, has the fidelity of $0.990$ with the Bell-state $|\Psi\rangle$. The figure shows the real part of the density matrix, for imaginary part and for the state preparation for reduced frequency correlations, see the Supplemental Material.
}
\end{figure}

In the experiment, we consider fibers of equal lengths $L_1=L_2=L$ with 
$\Delta n_1=\Delta n_2=\Delta n$.  In order to quantify the enhancement due to 
the nonlocal character of the channel, we take the ratio of lengths $L$ capable of 
distributing a certain given amount of entanglement $E$ for the 
channel with correlations $L^{K}(E)$, and without correlations $L^0(E)$. Thus, the 
enhancement factor $R(K)$ is given by
\begin{equation}
R(K):= \frac{L^{K}(E)}{L^0(E)}=\frac{1}{\sqrt{1-|K|}} =\sqrt{\frac{2\ln E(L,0)}{\ln E_{K}(L,L)}} .
\label{eq2}
\end{equation}
For the derivation of the last step, see the Supplemental Material. Here,  
$E(L,0)$ is the polarization entanglement using one fiber of length $L$ and $E_{K}(L,L)$ the entanglement using two fibers of length $L$ with the correlation coefficient $K$.
We also see that the enhancement factor lies between $R(0)=1$ for 
uncorrelated noise and $R(-1)\to\infty$ for maximally correlated noise. Note that
the correlation coefficient is assumed to be negative here, corresponding to the
initial Bell state $\ket{\Psi}$ \cite{nlnm}. If one considers instead the Bell states
$\frac{1}{\sqrt{2}}(\ket{HV}\pm\ket{VH})$ the correlation coefficient
must be taken to be positive.

\begin{figure}[t]
\centering
\includegraphics[width=0.4\textwidth]{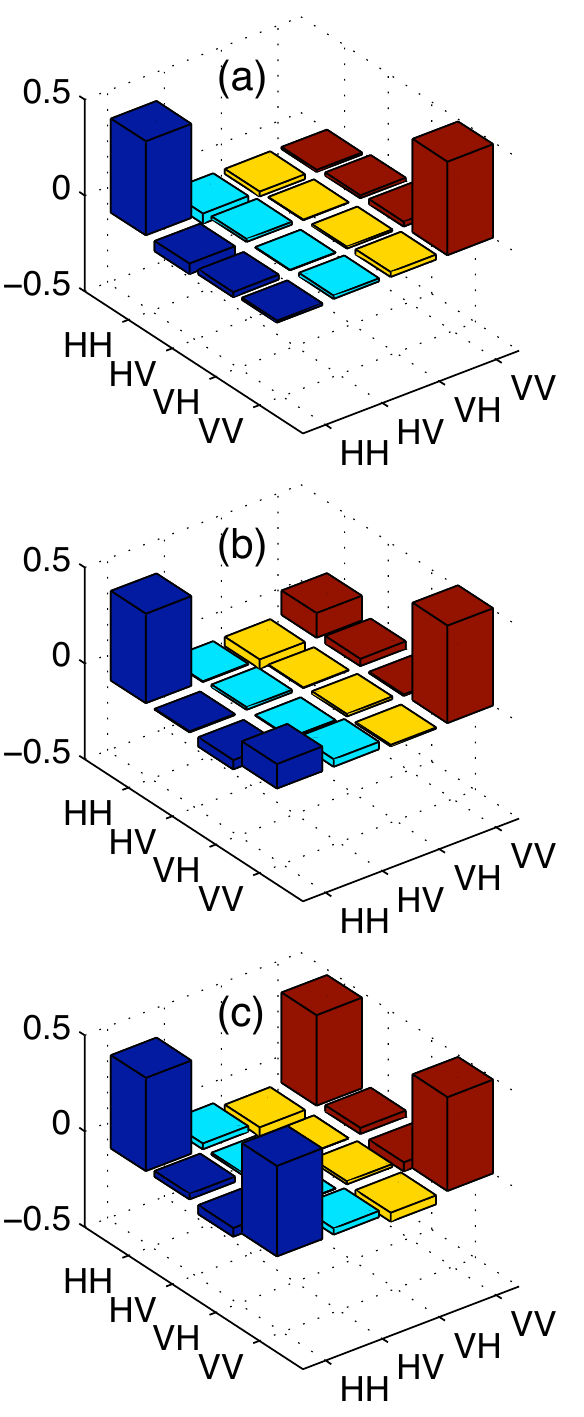}
\caption{\label{fig3} 
(color online) The polarization state tomography after the channels.
(a) One local channel (LOC) (b) intermediate nonlocal channels (NL1) (c) highly nonlocal channels (NL2).
Increasing the nonlocal character of the channels allows to protect the polarization entanglement against the local dephasing noise.}
\end{figure}

The experimental setup is presented in Fig.~1 and is divided into three parts: state preparation, PMD channel, and tomography. The state preparation generates the Bell state $\ket{\Psi}=\frac{1}{\sqrt{2}}(\ket{HH}+\ket{VV})$ by spontaneous parametric downconversion (SPDC) (for other states, see the Supplemental Material).  A $100 \rm{mW}$, V-polarized beam with a controllable spectral width at $351.1\rm{nm}$ from an Argon laser impinges on HWP0 whose optical axis is deviated $22.5^{\circ}$ from vertical direction. Then it pumps a pair of phase-matched type I $\beta$-barium borate (BBO) crystals whose optical axes are orthogonal to each other to prepare a pair of entangled photons~\cite{Kwiat93}. HWP2 and HWP3 rotate $H$ and $V$ to the fast and slow axes of the PM fiber. As the FWHM of the interference filter is $4 \rm{nm}$, much narrower than the width of the photons from SPDC, the photons locally have a band width of $4 \rm{nm}$. In the tomography part, HWP4 and HWP5 rotate the polarization directions of fast and slow axes of the PMF back to $H$ and $V$, followed by a standard tomography process~\cite{Jam01} with the determination of the entanglement (concurrence) of the state.

In the experiment, we study three different PMD channels: A local channel (LOC) where the single frequency mode of the pump light is used and only one of the photons transmits through a $100\rm{m}$ PM fiber, an intermediate nonlocal channel (NL1) where the single line mode of the pump light is used and both of the photons transmit through $100\rm{m}$ PM fibers, and a highly nonlocal channel (NL2) where also the single frequency mode of the pump light is used but both of the photons transmit through $100\rm{m}$ PM fibers. The spectral widths for the single line mode and the single frequency mode are about 20GHz and 10MHz, respectively. 
 
In Fig.~2 an example of experimentally measured initial density matrix is shown. The state is prepared using a narrow pump width resulting into high anticorrelation in the frequencies of the photons.  The polarization state has initial entanglement of $0.98$ and fidelity equal to $0.990$ with the Bell state $\ket{\Psi}$ mentioned above. Increasing the pump width uncorrelates the frequencies of the two photons, but leaves the initial polarization state almost intact (fidelity with $\ket{\Psi}$ is equal to 0.981). 

The photons are then sent through the PM fibers which realise, depending on the amount of anticorrelation in the frequencies of the photons, either the local channel (LOC), the intermediate nonlocal channel (NL1) or the highly nonlocal channel (NL2) on the initial polarization state. The results of the state tomography after the different channels are shown in Fig.~3. We can clearly see how the anticorrelation in the frequencies of the photons allows to protect the polarization entanglement: The local channel [Fig.~3 (a)] destroys the coherences and thus the entanglement of the two-photon polarization state entirely. A partial protection of the entanglement can be achieved, when the composite environment contains some correlations [Fig.~3 (b)]. When approaching perfect anticorrelation in the frequency [Fig.~3 (c)] the polarization entanglement of the initial Bell state $\ket{\Psi}$ is almost completely protected from local dephasing noise.

To further illustrate the enhancement in the entanglement distribution due to the environment correlations, Table~1 shows the numerical values of the fidelity, the entanglement, and the enhancement factor $R$ from Eq.~(7) for all of the cases considered above (for details see the Supplemental Material). We see that a nonlocal channel can clearly outperform a local one: even in the case of an initial multimode pump pulse with intermediate value for initial frequency correlations, the enhancement factor is $2$, i.e., one can double the distance over which entanglement can be distributed. By eliminating the longitudinal modes and constructing a highly nonlocal channel, we achieve an enhancement factor of $12$.
 
 \begin{table}[t]
\begin{tabular}{|c|c|c|c|c|c|}
\hline
& $F(0 \rm{m})$ &$E(0 \rm{m})$ & $F(100 \rm{m})$ & $E(100 \rm{m})$ & $R$ \\
\hline

LOC & 0.990 $\pm$ 0.005 & 0.98 $\pm$ 0.02 & 0.691 $\pm$ 0.005 & 0.04 $\pm$ 0.01 &  1 \\
\hline
NL1 & 0.981 $\pm$ 0.005 & 0.93 $\pm$ 0.02 & 0.78 $\pm$ 0.01& 0.24 $\pm$ 0.03 & 2 \\
\hline
NL2 & 0.990 $\pm$ 0.005 & 0.98 $\pm$ 0.02 & 0.979 $\pm$ 0.003 & 0.96 $\pm$ 0.01 &12\\
\hline
\end{tabular}
\caption{\label{tab1} The fidelities with the maximally entangled state [$F(L)$], the entanglement measured by concurrence [$E(L)$], and the enhancement factor ($R$) for the different channels. LOC stands for the local channel with a fiber in only one of the arms, NL1 for intermediate nonlocal channel, and NL2 for highly nonlocal channel. The measured CHSH value for the NL2 channel after the distribution is $2.66\pm 0.04$.
Note that the reported values of $R$ present the lower bounds since the concurrence is not exactly equal to one at the initial state preparation. For more details including the counting statistics and error bars, see the Supplemental Material.
}
\end{table}

In conclusion, we have experimentally demonstrated how initial correlations in the 
frequencies of the two photons and the subsequent nonlocal character of the 
channel enhance entanglement distribution in PM optical fibers. By 
enlarging the frequency correlations and nonlocal memory effects, the distances over which entanglement can 
be shared increase by a multiple with respect to the uncorrelated case.
In our experiment we are 
able to increase the entanglement distribution distance by a factor of 
at least 12, by tuning the spectral width of the original pump without affecting the 
initial polarization entanglement. It is important to notice that our scheme ultimately 
relies on techniques that are incorporated prior to the creation of the entangled 
photon pair and does not need any additional components after the pair has been 
created. This also means that our scheme can be incorporated in a rather 
straightforward manner in already existing schemes which use entanglement or 
polarization protection operations during the actual distribution process. 
\vspace{1cm}

\acknowledgments
 This work was supported by the National Basic Research Program of China (Grants No. 2011CB921200), the CAS, the National Natural Science Foundation of China (Grant Nos. 11274289, 11104261, 61108009 and 61222504), the Fundamental Research Funds for the Central Universities (No. WK2470000011), Anhui Provincial Natural Science Foundation (No. 1208085QA08),
the Academy of Finland (mobility from Finland 259827), the Jenny and Antti 
Wihuri Foundation (Finland), the Graduate School of Modern Optics and 
Photonics (Finland), and the German Academic Exchange Service (DAAD).

\end{document}